\documentclass[12pt]{article}
\pdfoutput=1

\usepackage{amsmath}
\usepackage{amssymb}
\usepackage{epsfig}
\usepackage{epstopdf}
\usepackage{latexsym}
\usepackage{color}
\numberwithin{equation}{section}
\usepackage{slashed}
\usepackage[
      colorlinks=false,
      linkcolor=darkblue,  
      urlcolor=blue,    
      filecolor=blue,     
      citecolor=red,
linktocpage=true,
      pdfstartview=FitV,
      bookmarksopen=true    
      ]{hyperref}

\usepackage{amssymb}

\begin{document}

\begin{titlepage}

\centerline
\centerline
\centerline
\bigskip
\bigskip
\centerline{\Huge \rm Equivariant localization for wrapped}
\bigskip
\centerline{\Huge \rm M5-branes and D4-branes}
\bigskip
\bigskip
\bigskip
\bigskip
\bigskip
\bigskip
\bigskip
\centerline{\rm Minwoo Suh}
\bigskip
\centerline{\it School of General Education, Kumoh National Institute of Technology,}
\centerline{\it Gumi, 39177, Korea}
\bigskip
\centerline{\tt minwoosuh1@gmail.com} 
\bigskip
\bigskip
\bigskip
\bigskip
\bigskip
\bigskip
\bigskip
\bigskip

\begin{abstract}
\noindent Following the work of \cite{BenettiGenolini:2024kyy}, we consider six- and seven-dimensional gauged supergravity coupled to a vector multiplet on an ansatz of $AdS_{4,5}\times{M}_2$, respectively, where $M_2$ is a two-dimensional surface with a Killing vector. We construct equivariantly closed forms from spinor bilinears of Killing spinors. From the integration of equivariantly closed forms via the Berline-Vergne-Atiyah-Bott fixed point formula, we derive the gravitational blocks which were previously conjectured. For the construction of equivariantly closed forms, we formulate six- and seven-dimensional gauged supergravity in the very special geometry. 
\end{abstract}

\vskip 6cm

\flushleft {April, 2024}

\end{titlepage}

\tableofcontents

\section{Introduction}

One of the lessons from the recent development in black hole microstate counting, \cite{Benini:2015eyy}, was that the Bekenstein-Hawking entropy can be obtained by extremizing an off-shell quantity known as the entropy function. This idea was turned out to be universal for any observables dual to partition functions of superconformal field theories in diverse dimensions by the AdS/CFT correspondence, \cite{Maldacena:1997re}. The expressions of these off-shell quantities, namely the``gravitational blocks", \cite{Hosseini:2019iad}, were conjectured and proved to reproduce the corresponding observables, $e.g.$, the Bekenstein-Hawking entropies, central charges, and free energies. Later, the gravitational blocks for solutions with orbifold singularities, $i.e.$, spindle, were proposed, \cite{Hosseini:2021fge, Faedo:2021nub}, and proved to calculate the duals of partition functions, \cite{Suh:2022olh, Faedo:2022rqx, Hosseini:2023ewi, Suh:2023xse, Hristov:2023rel, Faedo:2024upq}. See \cite{Boido:2022iye, Boido:2022mbe} also for an approach from string and M-theory.

Recently, for the supergravity solutions with an R-symmetry Killing vector, it was shown that observables dual to partition functions in field theory can be calculated by integrating equivariantly closed forms which are constructed by spinor bilinears of Killing spinors, \cite{BenettiGenolini:2023kxp, Martelli:2023oqk, BenettiGenolini:2023yfe, BenettiGenolini:2023ndb, Colombo:2023fhu, BenettiGenolini:2024kyy}. The Berline-Vergne-Atiyah-Bott fixed point formula, \cite{Berline:1982a, Atiyah:1984px}, was utilized to perform the integrations. In particular, in \cite{BenettiGenolini:2024kyy}, this method of equivariant localization was applied to gauged supergravity theories and served as a derivation of gravitational blocks for four- and five-dimensional gauged supergravity. In this work, we apply the method of equivariant localization to six- and seven-dimensional gauged supergravity and derive the gravitational blocks for corresponding theories in \cite{Faedo:2021nub}.

In \cite{BenettiGenolini:2024kyy}, in order to construct the equivariantly closed forms from spinor bilinears of Killing spinors, the structure of respective gauged supergravity theory was exploited, $e.g.$, the very special geometry of five-dimensional gauged supergravity, \cite{deWit:1992cr}, and the $\mathcal{N}=2$ formalism of four-dimensional gauged supergravity. As a foundation of construction, we look out for structures and reveal the structure of very special geometry in six- and seven-dimensional guaged supergravity coupled to a vector multiplet. To the best of our knowledge, it is the first time to formulate six- and seven-dimensional gauged supergravity in the very special geometry. Utilizing the structure of very special geometry, we construct the equivariantly closed forms for the general ansatz of $AdS_5\times{M}_2$ and $AdS_4\times{M}_2$ where a Killing vector is assumed to be on a two-dimensional surface, $M_2$. We derive the gravitational blocks proposed in \cite{Faedo:2021nub} by performing the integrations of equivariantly closed forms via the Berline-Vergne-Atiyah-Bott fixed point formula. To be specific, we study the cases of $M_2$ to be a Riemann surface with topological twist and a spindle with two classes of twists.

We closely follow the presentation of \cite{BenettiGenolini:2024kyy} as it is well organized and to promote comparison of the studies in diverse dimensions.

In section \ref{sectwo}, we review the equivariant localization formula of Berline-Vergne-Atiyah-Bott. In section \ref{sec2}, we construct the equivariantly closed forms and derive the gravitational blocks for M5-branes on a two-dimensional surface, $M_2$. In section \ref{sec3}, we construct the equivariantly closed forms and derive the gravitational blocks for D4-branes on a two-dimensional surface, $M_2$. We conclude with some outlook in section \ref{sec4}. The construction of spinor bilinears are presented in appendix \ref{appA} and \ref{appB}.

\section{Equivariant localization} \label{sectwo}

In this section, we review the equivariant localization formula of Berline-Vergne-Atiyah-Bott (BVAB), \cite{Berline:1982a, Atiyah:1984px}, as it was formulated in \cite{BenettiGenolini:2023kxp, BenettiGenolini:2023ndb}.

We consider compact Riemannian manifolds, $(M,g)$, of dimension $2n$, with a $U(1)$ Killing vector, $\xi$. For the polyforms, which is a formal sum of different forms of various degrees, we introduce the equivariant differential, 
\begin{equation}
d_\xi\,\equiv\,d-\xi\lrcorner\,,
\end{equation}
which squares to minus the Lie derivative along $\xi$: $d_\xi^2\,=\,-\mathcal{L}_\xi$. Then, for the polyforms, $\Phi$, which is invariant under the equivariant differential, $\mathcal{L}_\xi\Phi=0$, the differential is nilpotent and we can construct an equivariant cohomology. We consider the case that there is a non-trivial fixed point set $F\equiv\{\xi=0\}\subset{M}$. The integral of a polyform is defined to be the integral of its top component.

For a connected component of $F$ with codimension $2k$, $\xi$ generates a linear isometric action on the rank-$2k$ normal bundle. Then we have
\begin{equation}
\xi\,=\,\sum_{i=1}^k\epsilon_i\partial_{\varphi_i}\,,
\end{equation}
where $\varphi_i$ are polar angles with period, $\Delta\varphi_i=2\pi$, on the $i$th two-plane, $\mathbb{R}^2\subset\mathbb{R}^{2k}$, in a normal space. For generic $\epsilon_i$, the normal bundle, $\mathcal{N}$, to $F$ splits into a sum of complex line bundles, $\mathcal{N}\cong{L}_1\oplus\cdots\oplus{L}_k$, with $L_i$ rotated by $\partial_{\varphi_i}$.

The equivariant localization formula gives the integral of an equivariantly closed form by sum of connected components of the fixed point set of various dimensions, \cite{BenettiGenolini:2023kxp, BenettiGenolini:2023ndb},
\begin{align}
\int_M\Phi\,=\,\sum_{\text{dim}\,0}\frac{1}{d_{F_0}}\frac{(2\pi)^n}{\epsilon_1\cdots\epsilon_n}\Phi_0+&\sum_{\text{dim}\,2}\frac{1}{d_{F_2}}\frac{(2\pi)^{n-1}}{\epsilon_1\cdots\epsilon_{n-1}}\int\Big[\Phi_2-\Phi_0\sum_{1\le{i}\le{n}-1}\frac{2\pi}{\epsilon_i}c_1(L_i)\Big] \notag \\
+\sum_{\text{dim}\,4}\frac{1}{d_{F_0}}\frac{(2\pi)^{n-2}}{\epsilon_1\cdots\epsilon_{n-2}}\int&\Big[\Phi_4-\Phi_2\wedge\sum_{1\le{i}\le{n}-2}\frac{2\pi}{\epsilon_i}c_1(L_i) \notag \\
&+\Phi_0\sum_{1\le{i}\le{j}\le{n}-2}\frac{(2\pi)^2}{\epsilon_i\epsilon_j}c_1(L_i)\wedge{c}_1(L_j)\Big]+\cdots\,,
\end{align}
where the normal space to a subspace, $F_{2n-2k}$, is $\mathbb{R}^{2k}/\Gamma$, where $\Gamma\subset{SO}(2k)$ is the finite group of order $d_{F_{2n-2k}}$ defining the orbifold structure. 

\section{M5-branes wrapped on $M_2$} \label{sec2}

\subsection{$U(1)^2$-gauged supergravity in seven dimensions}

We consider $U(1)^2$-gauged supergravity in seven dimensions, \cite{Liu:1999ai}, which is a consistent truncation of maximal gauged supergravity in seven dimensions, \cite{Pernici:1984xx}. 

In order to study equivariant localization, it is convenient to use the geometric structure of model, \cite{BenettiGenolini:2024kyy}. In this section, we present $U(1)^2$-gauged supergravity in seven dimensions in the structure of {\it very special geometry}, \cite{deWit:1992cr}.

The bosonic action is given by{\footnote{We rescaled the gauge fields in \cite{Liu:1999ai} by $F^\text{there}\rightarrow\frac{1}{2}F^\text{here}$ to have the standard normalization of Maxwell action.}}
\begin{equation} \label{sixtheory7}
S_7\,=\,\frac{1}{16\pi{G}_N^{(7)}}\int_{M_7}\left(R_7-\mathcal{V}-\mathcal{G}_{ij}\partial_\mu\varphi^i\partial^\mu\varphi^j-\frac{1}{2}G_{IJ}F_{\mu\nu}^IF^{J\mu\nu}\right)\text{vol}_7\,,
\end{equation}
in the mostly plus signature. $R_7$ is the Ricci scalar and $\text{vol}_7$ is the volume form. We have two real scalar fields, $\varphi^1$ and $\varphi^2$. We introduce the following parametrization of the scalar fields,
\begin{equation}
X^I\,=\,\left(e^{\frac{4}{\sqrt{10}}\varphi_2},e^{-\frac{1}{\sqrt{2}}\varphi_1-\frac{1}{\sqrt{10}}\varphi_2},e^{\frac{1}{\sqrt{2}}\varphi_1-\frac{1}{\sqrt{10}}\varphi_2}\right)\,.
\end{equation}
There are two $U(1)$ gauge fields, $A^1$ and $A^2$, and we have
\begin{equation} \label{gaugefd7}
A^I\,=\,\left(A^0\equiv0,A^1,A^2\right)\,,
\end{equation}
where, in order to formulate the very special geometry, we have introduced a trivial gauge field, $A^0$. The metric for the kinetic terms of scalar fields and gauge fields are given, respectively, by
\begin{equation}
\mathcal{G}_{ij}\,=\,\frac{1}{2}\text{diag}\left(1,1\right)\,,
\end{equation}
and
\begin{equation}
G_{IJ}\,=\,\text{diag}\left(\frac{1}{4\left(X^0\right)^2},\frac{1}{2\left(X^1\right)^2},\frac{1}{2\left(X^2\right)^2}\right)\,.
\end{equation}
The scalar potential is given by
\begin{equation}
\mathcal{V}\,\equiv\,\xi_I\xi_J\left(\mathcal{G}^{ij}\partial_iX^I\partial_jX^J-\frac{6}{5}X^IX^J\right)\,,
\end{equation}
where $\partial_I\equiv\partial_{X^I}$ and $\partial_i\equiv\partial_{\varphi^i}$. We have the Fayet-Iliopoulos (FI) gauging parameters,
\begin{equation}
\xi_I\,=\,\frac{1}{2}\left(m,g,g\right)\,,
\end{equation}
where $g$ and $m$ are gauge coupling constant and mass parameter, respectively. In order to have the supersymmetric vacuum, it is required to be $g=2m$.

The expressions are neatly packaged by introducing a prepotential which satisfies the constraint,
\begin{equation} \label{prepotential7}
\mathcal{F}\left(X^I\right)\,\equiv\,X^0\left(X^1X^2\right)^2\,=\,1\,.
\end{equation}
The metrics are obtained from
\begin{equation}
G_{IJ}\,=\,-\frac{1}{4}\partial_I\partial_J\log\mathcal{F}\Big|_{\mathcal{F}=1}\,, \qquad \mathcal{G}_{ij}\,=\,\partial_iX^I\partial_jX^JG_{IJ}\Big|_{\mathcal{F}=1}\,,
\end{equation}
and the indices can be raised and lowered by
\begin{equation}
X_I\,\equiv\,G_{IJ}X^J\,, \qquad \partial_iX_I\,=\,-G_{IJ}\partial_iX^J\,.
\end{equation}
We also find an expression, 
\begin{equation} \label{GIJGIJ7}
G^{IJ}\,=\,\mathcal{G}^{ij}\partial_iX^I\partial_jX^J+\frac{4}{5}X^IX^J\,,
\end{equation}
for the inverse of $G_{IJ}$. Furthermore, we have $X^IX_I=5/4$.

The supersymmetry variations of the gravitino and gaugino fields are given, respectively, by
\begin{align} \label{susyvar117}
0\,=&\,\left[\nabla_\mu-\frac{i}{2}\xi_IA^I_\mu+\frac{1}{10}W\Gamma_\mu+\frac{i}{20}X_IF^I_{\nu\rho}\left(\Gamma_\mu\,^{\nu\rho}-8\delta_\mu^\nu\Gamma^\rho\right)\right]\epsilon\,, \notag \\
0\,=&\,\left[-\frac{i}{2}\mathcal{G}_{ij}\partial_\mu\varphi^j\Gamma^\mu+\frac{i}{2}\partial_iW+\frac{1}{4}\partial_iX_IF^I_{\mu\nu}\Gamma^{\mu\nu}\right]\epsilon\,,
\end{align}
where $\Gamma_\mu$ are Cliff(1,6) generators. The superpotential is given by
\begin{equation}
W\,\equiv\,\xi_IX^I\,,
\end{equation}
and the scalar potential can be expressed in terms of the superpotential,
\begin{equation} \label{potsup7}
\mathcal{V}\,=\,\mathcal{G}^{ij}\partial_iW\partial_jW-\frac{6}{5}W^2\,.
\end{equation}
The supersymmetric $AdS_7$ vacuum of the scalar potential is dual to six-dimensional $\mathcal{N}=(2,0)$ superconformal field theories.

\subsection{The $AdS_5$ ansatz}

We consider the following background, \cite{BenettiGenolini:2024kyy},
\begin{equation} \label{ads4m2met7}
ds_7^2\,=\,e^{2\lambda}\left[ds_{AdS_5}^2+ds_{M_2}^2\right]\,,
\end{equation}
where $ds^2_{AdS_5}$ is a metric on $AdS_5$ of unit radius.  The scalar fields and $\lambda$ are functions on $M_2$ and $A^I$ are gauge fields on $M_2$. 

We dimensionally reduce the equations of motion from the action in \eqref{sixtheory7} on the background, \eqref{ads4m2met7}, and find that the reduced equations of motion can be obtained from the variations of following two-dimensional action, $S_7=\frac{\text{vol}_{AdS_5}}{16\pi{G}_N^{(7)}}S_2$,
\begin{align} \label{2daction7}
S_2\,=\,\left.\int_{M_2}\right[&e^{5\lambda}\left(R-20+30\left(\nabla\lambda\right)^2-\mathcal{G}_{ij}\partial_\mu\varphi^i\partial^\mu\varphi^j\right)-e^{7\lambda}\mathcal{V} \notag \\ 
&-\left.\frac{1}{2}e^{3\lambda}G_{IJ}F_{\mu\nu}^IF^{J\mu\nu}\right]\text{vol}\,,
\end{align}
where $R$, $\nabla$ and vol are the Ricci scalar, Levi-Civita connection and volume form on $M_2$. The Maxwell equations are
\begin{equation} \label{maxwell7}
d\left[e^{3\lambda}G_{IJ}\left(*F^J\right)\right]\,=\,0 \qquad \Longrightarrow \qquad d\left[e^{3\lambda}G_{IJ}F_{12}^J\right]\,=\,0\,, \qquad \forall I\,=\,0,1,2\,,
\end{equation}
where we have $F^I\,=\,F^I_{12}\text{vol}$ with $F_{12}^I$ a function on $M_2$. For the warp factor, $\lambda$, the equation of motion is
\begin{align} \label{warpeom7}
e^{5\lambda}\left[R-20+30\left(\nabla\lambda\right)^2-\mathcal{G}_{ij}\partial_\mu\varphi^i\partial^\mu\varphi^j\right]\,=&\,\frac{7}{5}e^{7\lambda}\mathcal{V}+\frac{3}{10}e^{3\lambda}G_{IJ}F_{\mu\nu}^IF^{J\mu\nu} \notag \\
&+\text{total\,\,derivative}\,.
\end{align}
The trace of Einstein equation is
\begin{equation} \label{traceeinstein7}
e^{7\lambda}\mathcal{V}-\frac{1}{2}e^{3\lambda}G_{IJ}F_{\mu\nu}^{I}F^{J\mu\nu}\,=\,-20e^{5\lambda}+\text{total\,\,derivative}\,.
\end{equation}

From \eqref{2daction7} the partially off-shell (POS) action is obtained by imposing \eqref{warpeom7},
\begin{equation}
S_2|_\text{POS}\,=\,\frac{2}{5}\int_{M_2}\left(e^{7\lambda}\mathcal{V}\text{vol}-e^{3\lambda}G_{IJ}F_{12}^IF^J\right)\,.
\end{equation}
Further imposing \eqref{traceeinstein7} we find the on-shell action,
\begin{equation}
S_2|_\text{OS}\,=\,-8\int_{M_2}e^{5\lambda}\text{vol}\,.
\end{equation}

\subsection{Equivariantly closed forms}

We introduce the Killing spinor on the background, \eqref{ads4m2met7}, for the supersymmetry variations, \eqref{susyvar117}, \cite{BenettiGenolini:2024kyy},
\begin{equation}
\epsilon\,=\,\vartheta\otimes{e}^{\lambda/2}\zeta\,,
\end{equation}
where $\vartheta$ is a Killing spinor on $AdS_5$, $\zeta$ is a spinor on $M_2$ and the warp factor is introduced for convenience. Then, as presented in appendix \ref{appA}, we obtain the supersymmetry equations, \eqref{kse117}.

We introduce the spinor bilinears in $\zeta$ on $M_2$, \cite{BenettiGenolini:2024kyy},
\begin{equation} \label{sbil7}
S\,=\,\zeta^\dagger\zeta\,, \quad P\,=\,\zeta^\dagger\gamma_3\zeta\,, \quad K\,=\,\zeta^\dagger\gamma_{(1)}\zeta\,, \quad \xi^\flat\,=\,-i\zeta^\dagger\gamma_{(1)}\gamma_3\zeta\,,
\end{equation}
where we have $\gamma_{(n)}=\frac{1}{n!}\gamma_{\mu_1\cdots\mu_n}dx^{\mu_1}\wedge\cdots\wedge{d}x^{\mu_n}$ and the chirality operator on $M_2$ is $\gamma_3=-i\gamma_1\gamma_2$. We obtain the spinor bilinear equations in \eqref{sbileq7}. We find that the Killing vector, $\xi$, on $M_2$ is dual to the one-form, $\xi^\flat$, and we have
\begin{equation}
d\xi^\flat\,=\,-2\left(4+PS^{-1}e^\lambda{W}\right)P\text{vol}\,.
\end{equation}

We found equivariantly closed forms under $d_\xi=d-\xi\mathbin{\lrcorner}$ in appendix \ref{appA}, \cite{BenettiGenolini:2024kyy},
\begin{equation} \label{phiF7}
\Phi^{F^I}\,\equiv\,F^I-X^Ie^\lambda{P}\,,
\end{equation}
and
\begin{equation}
\Phi^\text{vol}\,\equiv\,e^{7\lambda}\mathcal{V}\text{vol}-e^{6\lambda}WS\,.
\end{equation}
By employing the Maxwell equations in \eqref{maxwell}, we obtain another equivariantly closed form,
\begin{align} \label{phiS7}
\Phi^S\,\equiv&\,\frac{2}{5}\left(\Phi^\text{vol}-e^{3\lambda}{G}_{IJ}F^I_{12}\Phi^{F^J}\right) \notag \\
=&\,\frac{2}{5}\left(e^{7\lambda}\mathcal{V}\text{vol}-e^{3\lambda}G_{IJ}F^I_{12}F^J-e^{6\lambda}WS+e^{4\lambda}F^I_{12}X^JP\right)\,,
\end{align}
where the two-forms are the two-dimensional action in \eqref{2daction7}.

As we note that the scalar bilinear, $S$, is a constant in appendix \ref{appA}, we choose the normalization to be $S\equiv1$.

\subsection{Localization}

For the spinor bilinears in \eqref{sbil7}, we find $||\xi||^2=||K||^2=S^2-P^2$. At a fixed point, $p$, where we have $\xi|_p=0$, we obtain $P|_p=\pm{S}|_p=\pm1$, and the sign is the chirality of the spinor, $\zeta$, at the fixed point, $p$. For $\xi\ne0$, around the fixed point, $p$, we may have $M_2$ to be $\mathbb{R}^2/\mathbb{Z}_{d_p}$ with orbifold singularities and $\xi=\epsilon_p\partial_{\phi_p}$ where $\phi_p$ is a polar coordinate of period, $2\pi/d_p$. Let us name $\epsilon_p$ to be the weight of $\xi$ at the fixed point, $p$, \cite{BenettiGenolini:2024kyy}.

Now we employ the Berline-Vergne-Atiyah-Bott fixed point formula, \cite{Berline:1982a, Atiyah:1984px}, to perform the integrations of equivariantly closed forms in \eqref{phiF7} and \eqref{phiS7} on $M_2$, \cite{BenettiGenolini:2024kyy}. The magnetic charges are given by
\begin{equation} \label{magch7}
\mathfrak{p}^I\,\equiv\,\int_{M_2}\frac{F^I}{2\pi}\,=\,\int_{M_2}\frac{\Phi^{F^I}}{2\pi}\,=\,-\frac{1}{2\pi}\sum_{\text{fixed}\,p}\frac{1}{d_p}\frac{2\pi}{\epsilon_p}\left(X^Ie^\lambda{P}\right)\Big|_p\,,
\end{equation}
and the two-dimensional action is 
\begin{align} \label{twoac7}
S_2|_\text{POS}\,=\,\int_{M_2}\Phi^S\,=&\,-\frac{2}{5}\sum_{\text{fixed}\,p}\frac{1}{d_p}\frac{2\pi}{\epsilon_p}\left(e^{6\lambda}WS-e^{4\lambda}G_{IJ}F_{12}^IX^JP\right)\Big|_p \notag \\
=&\,2\sum_{\text{fixed}\,p}\frac{1}{d_p}\frac{2\pi}{\epsilon_p}e^{5\lambda}P\Big|_p\,,
\end{align}
where we employed the algebraic constraint in \eqref{algecon7} in the last line and $P^2|_p=S^2|_p=1$. As we have $d\xi^\flat|_p=2\epsilon_p\text{vol}$ at a fixed point, we find
\begin{equation} \label{dxidxi7}
e^\lambda{W}P|_p\,=\,-\left(4+\epsilon_pP_p\right)\,,
\end{equation}
where we used $P_p\,=\,\pm1$.

We introduce the new scalar fields, 
\begin{equation} \label{newsc7}
x^I\,\equiv\,-X^Ie^\lambda{P} \qquad \Longrightarrow \qquad \Phi^{F^I}\,=\,F^I+x^I\,.
\end{equation}
With $\mathcal{F}\left(X^I\right)=1$ in \eqref{prepotential7}, we obtain
\begin{equation}
\mathcal{F}\left(x^I\right)|_p\,=\,x^0\left(x^1x^2\right)^2|_p\,=\,-e^{5\lambda}P|_p\,,
\end{equation}
where we employed $P^2|_p=1$. Then \eqref{twoac7}, \eqref{magch7} and \eqref{dxidxi7} are given, respectively, by
\begin{equation} \label{constraint117}
S_2|_\text{POS}\,=\,-2\sum_{\text{fixed}\,p}\frac{2\pi}{d_p\epsilon_p}\mathcal{F}\left(x_p^I\right)\,, \qquad \mathfrak{p}^I\,=\,\sum_{\text{fixed}\,p}\frac{1}{d_p\epsilon_p}x_p^I\,, \qquad \xi_Ix_p^I\,=\,4+\epsilon_pP_p\,,
\end{equation}
where we introduced $x_p^I\equiv{x}^I|_p$.

Topologically a spindle is the only choice for $M_2$ if it is compact without boundary and has a $U(1)$ isometry with fixed points, \cite{Ferrero:2021wvk}. The spindle has two fixed points, $p=\pm$, and we choose $d_\pm=n_\pm$ with $n_\pm\ge1$. The Killing vector field on the spindle is given by
\begin{equation}
\xi\,=\,b_0\partial_\varphi\,,
\end{equation}
where $\varphi$ is an azimuthal coordinate on the spindle with period of $2\pi$. We can write $\epsilon_+=-b_0/n_+$, $\epsilon_-=b_0/n_-$. $P_\pm$ is arbitrary and $|P_\pm|=1$. Then the constraints in \eqref{constraint117} are given by
\begin{equation} \label{constraints227}
\mathfrak{p}^I\,=\,-\frac{1}{b_0}\left(x_+^I-x_-^I\right)\,, \qquad \xi_Ix_+^I\,=\,4-\frac{b_0}{n_+}P_+\,, \qquad \xi_Ix_-^I\,=\,4+\frac{b_0}{n_-}P_-\,,
\end{equation}
and, from these, we also obtain
\begin{equation} \label{magneticcon7}
\xi_I\mathfrak{p}^I\,=\,\frac{P_+}{n_+}+\frac{P_-}{n_-}\,=\,\frac{n_-P_++n_+P_-}{n_-n_+}\,.
\end{equation}
If we introduce $\sigma\equiv{P}_+/P_-$, $\sigma=+1$ is for the case of equal chirality of spinor at two poles of spindle, namely twist. $\sigma=-1$ is for the opposite chiralities, namely anti-twist, \cite{Ferrero:2021etw}. The off-shell action is given by
\begin{equation} \label{S2pos227}
S_2|_\text{POS}\,=\,\frac{4\pi}{b_0}\left[\mathcal{F}\left(x_+^I\right)-\mathcal{F}\left(x_-^I\right)\right]\,.
\end{equation}


\subsection{The central charge}

We introduce a new parametrization for the scalar fields with magnetic charges,
\begin{equation} \label{phi0127}
\phi_0\,\equiv\,\frac{m}{2}\left(x_\pm^0\pm\mathfrak{p}^0\frac{b_0}{2}\right)\,, \qquad \phi_1\,\equiv\,\frac{g}{2}\left(x_\pm^1\pm\mathfrak{p}^1\frac{b_0}{2}\right)\,, \qquad \phi_2\,\equiv\,\frac{g}{2}\left(x_\pm^2\pm\mathfrak{p}^2\frac{b_0}{2}\right)\,.
\end{equation}
As the gauge field, $A^0$, is identically zero, \eqref{gaugefd7}, we have $\mathfrak{p}^0=0$. Accordingly, we choose $\phi_0=2$. Namely, we have
\begin{equation}
\phi_0\,=\,2\,, \qquad \mathfrak{p}^0\,=\,0\,.
\end{equation}
For the choice of $\phi_0=2$, from \eqref{phi0127} and \eqref{newsc7}, we find 
\begin{equation}
2\,=\,\frac{m}{2}\left(-X^0e^\lambda{P}\right)|_\pm\,.
\end{equation}
As $X^0>0$ and $e^\lambda>0$, only the solutions in the twist class are allowed, $P_\pm=-1$, and it is consistent with \cite{Ferrero:2021etw}. We further introduce new parameters for the rest of magnetic charges and $b_0$,
\begin{equation}
\epsilon\,\equiv\,\frac{b_0}{2}\,, \qquad \mathfrak{n}_1\,\equiv\,\frac{g}{2}\mathfrak{p}^1\,, \qquad \mathfrak{n}_2\,\equiv\,\frac{g}{2}\mathfrak{p}^2\,.
\end{equation}
Then, from the constraints in \eqref{constraints227}, using the new parameters, we find 
\begin{align} \label{gbcon7}
\mathfrak{n}_1+\mathfrak{n}_2\,=&\,\frac{n_++\sigma{n}_-}{n_-n_+}\,, \notag \\
\phi_1+\phi_2-\frac{n_+-\sigma{n}_-}{n_-n_+}\epsilon\,=&\,2\,,
\end{align}
where we keep $\sigma\equiv{P}_+/P_-$ arbitrary. These are precisely the constraints on the magnetic charges and the variables, $\phi_i$, for M5-branes wrapped on a spindle in section 5.4 of \cite{Faedo:2021nub}.{\footnote{In \cite{Faedo:2021nub} $\phi_i$ was denoted by $\varphi_i$.}} The prepotential is then given by
\begin{equation}
\mathcal{F}\left(x_\pm^I\right)\,=\,\frac{4^3}{g^4m}\Big(\left(\phi_1\mp\mathfrak{n}_1\epsilon\right)\left(\phi_2\mp\mathfrak{n}_2\epsilon\right)\Big)^2\,.
\end{equation}
From \eqref{S2pos227}, the two-dimensional off-shell action is 
\begin{equation}
S_2|_\text{POS}\,=\,-\frac{2\pi}{\epsilon}\frac{4^3}{g^4m}\left[\Big(\left(\phi_1+\mathfrak{n}_1\epsilon\right)\left(\phi_2+\mathfrak{n}_2\epsilon\right)\Big)^2-\Big(\left(\phi_1-\mathfrak{n}_1\epsilon\right)\left(\phi_2-\mathfrak{n}_2\epsilon\right)\Big)^2\right]\,.
\end{equation}
From this, we recover the gravitational block in \cite{Faedo:2021nub},
\begin{equation} \label{offfree7}
F^-\left(\phi_i,\epsilon;\mathfrak{n}_i\right)\,=\,\frac{9g^4m}{2\,4^7\pi}N^3S_2|_\text{POS}\,.
\end{equation}
By extremizing the gravitational block or the off-shell action subject to the constraints, \eqref{gbcon7}, we find, \cite{Faedo:2021nub},
\begin{align} \label{phistar7}
\phi_1^*\,=&\,1+\frac{\left(\mathtt{s}+\mathfrak{n}_1+\mathfrak{n}_2\right)\left[2n_+^2n_-^2\left(\mathfrak{n}_1^2-\mathfrak{n}_2^2\right)+3\left(n_+-\sigma{n}_-\right)^2+n_+^2n_-^2\left(\mathfrak{n}_1-\mathfrak{n}_2\right)\mathtt{s}\right]}{12n_+n_-\left(\sigma-n_+n_-\mathfrak{n}_1\mathfrak{n}_2\right)\left[\mathtt{s}+2\left(\mathfrak{n}_1+\mathfrak{n}_2\right)\right]}\,, \notag \\
\epsilon^*\,=&\,\frac{\left(n_+-\sigma{n}_-\right)\left(\mathtt{s}+\mathfrak{n}_1+\mathfrak{n}_2\right)}{2\left(\sigma-n_+n_-\mathfrak{n}_1\mathfrak{n}_2\right)\left[\mathtt{s}+2\left(\mathfrak{n}_1+\mathfrak{n}_2\right)\right]}\,,
\end{align}
where we have defined
\begin{equation}
\mathtt{s}\,\equiv\,\sqrt{7\left(\mathfrak{n}_1^2+\mathfrak{n}_2^2\right)+2\mathfrak{n}_1\mathfrak{n}_2-6\frac{n_+^2+n_-^2}{n_+^2n_-^2}}\,.
\end{equation}
Inserting $\phi_1^*$ and $\epsilon^*$ in \eqref{offfree7}, we obtain the central charge, $a$, of dual four-dimensional superconformal field theories, \cite{Ferrero:2021wvk, Faedo:2021nub},
\begin{equation}
F^-\left(\phi_i^*,\epsilon^*;\mathfrak{n}_i\right)\,=\,\frac{3N^3}{8}\frac{\mathfrak{n}_1^2\mathfrak{n}_2^2\left(\mathtt{s}+\mathfrak{n}_1+\mathfrak{n}_2\right)}{\left(\frac{\sigma}{n_+n_-}-\mathfrak{n}_1\mathfrak{n}_2\right)\left[\mathtt{s}+2\left(\mathfrak{n}_1+\mathfrak{n}_2\right)\right]^2}\,,
\end{equation}
for the solutions in the twist class, $\sigma=+1$.

By plugging $x_-^I=x_+^I+b_0\mathfrak{p}^I$ from \eqref{constraints227} into \eqref{S2pos227} with $\sigma=+1$, we note that
\begin{equation} \label{spherelim7}
\lim_{b_0\rightarrow0}S_2|_\text{POS}\,=\,-4\pi\mathfrak{p}^I\frac{\partial}{\partial{x}^I}\mathcal{F}\left(x^I\right)\,.
\end{equation}
We have $x_-^I=x_+^I=x^I$ to be constant on $M_2$. From \eqref{magneticcon7} we find $\xi_Ix^I=2$. We can extremize \eqref{spherelim7} with the constraint, $\xi_Ix^I=2$. In particular, for $n_+=n_-=1$, as already noticed in \cite{Faedo:2021nub}, we should find the central charge obtained in \cite{Maldacena:2000mw, Bah:2011vv, Bah:2012dg} for $AdS_5\times\Sigma_\mathfrak{g}$ solutions with a topological twist on the Riemann surface of genus, $\mathfrak{g}$, and we have $\xi_I\mathfrak{p}^I=2P_+$ and $\sigma=+1$.

\section{D4-branes wrapped on $M_2$} \label{sec3}

\subsection{$U(1)^2$-gauged supergravity in six dimensions}

We consider $F(4)$ gauged supergravity, \cite{Romans:1985tw}, coupled to a vector multiplet, \cite{Andrianopoli:2001rs}, in six dimensions. The model was first presented in \cite{Karndumri:2015eta} and the action was explicitly given in \cite{Suh:2018szn}. For the parametrizations and conventions of the model, we follow \cite{Faedo:2021nub}.

In order to study equivariant localization, it is convenient to use the geometric structure of the model, \cite{BenettiGenolini:2024kyy}. In this section, we present $U(1)^2$-gauged supergravity in six dimensions in the structure of {\it very special geometry}, \cite{deWit:1992cr}.

The bosonic action is given by
\begin{equation} \label{sixtheory}
S_6\,=\,\frac{1}{16\pi{G}_N^{(6)}}\int_{M_6}\left(R_6-\mathcal{V}-\mathcal{G}_{ij}\partial_\mu\varphi^i\partial^\mu\varphi^j-\frac{1}{2}G_{IJ}F_{\mu\nu}^IF^{J\mu\nu}\right)\text{vol}_6\,,
\end{equation}
in the mostly plus signature. $R_6$ is the Ricci scalar and $\text{vol}_6$ is the volume form. We have two real scalar fields, $\sigma$ and $\varphi_2$,
\begin{equation}
\varphi^i\,=\,\left(\varphi_1\equiv\sigma,\varphi_2\right)\,.
\end{equation}
We introduce the following parametrization of the scalar fields,
\begin{equation}
X^I\,=\,\left(e^{-3\sigma},e^{\sigma+\varphi_2},e^{\sigma-\varphi_2}\right)\,.
\end{equation}
There are two $U(1)$ gauge fields, $A^1$ and $A^2$, and we have
\begin{equation} \label{gaugefd}
A^I\,=\,\left(A^0\equiv0,A^1,A^2\right)\,,
\end{equation}
where, in order to formulate the very special geometry, we have introduced a trivial gauge field, $A^0$. The metric for the kinetic terms of scalar fields and gauge fields are given, respectively, by
\begin{equation}
\mathcal{G}_{ij}\,=\,\text{diag}\left(4,1\right)\,,
\end{equation}
and
\begin{equation}
G_{IJ}\,=\,\text{diag}\left(\frac{1}{3\left(X^0\right)^2},\frac{1}{2\left(X^1\right)^2},\frac{1}{2\left(X^2\right)^2}\right)\,.
\end{equation}
The scalar potential is given by
\begin{equation}
\mathcal{V}\,\equiv\,\xi_I\xi_J\left(\mathcal{G}^{ij}\partial_iX^I\partial_jX^J-\frac{5}{4}X^IX^J\right)\,,
\end{equation}
where $\partial_I\equiv\partial_{X^I}$ and $\partial_i\equiv\partial_{\varphi^i}$. We have the Fayet-Iliopoulos (FI) gauging parameters,
\begin{equation}
\xi_I\,=\,\left(m,g,g\right)\,,
\end{equation}
where $g$ and $m$ are gauge coupling constant and mass parameter, respectively. In order to have the supersymmetric vacuum, it is required to be $g=3m$.

The expressions are neatly packaged by introducing a prepotential which satisfies the constraint,
\begin{equation} \label{prepotential}
\mathcal{F}\left(X^I\right)\,\equiv\,X^0\left(X^1X^2\right)^{3/2}\,=\,1\,.
\end{equation}
The metrics are obtained from
\begin{equation}
G_{IJ}\,=\,-\frac{1}{3}\partial_I\partial_J\log\mathcal{F}\Big|_{\mathcal{F}=1}\,, \qquad \mathcal{G}_{ij}\,=\,\partial_iX^I\partial_jX^JG_{IJ}\Big|_{\mathcal{F}=1}\,,
\end{equation}
and the indices can be raised and lowered by
\begin{equation}
X_I\,\equiv\,G_{IJ}X^J\,, \qquad \partial_iX_I\,=\,-G_{IJ}\partial_iX^J\,.
\end{equation}
We also find an expression, 
\begin{equation} \label{GIJGIJ}
G^{IJ}\,=\,\mathcal{G}^{ij}\partial_iX^I\partial_jX^J+\frac{3}{4}X^IX^J\,,
\end{equation}
for the inverse of $G_{IJ}$. Furthermore, we have $X^IX_I=4/3$.

The supersymmetry variations of the gravitino and gaugino fields are given, respectively, by
\begin{align} \label{susyvar11}
0\,=&\,\left[\nabla_\mu-\frac{i}{2}\xi_IA^I_\mu+\frac{1}{8}W\Gamma_\mu+\frac{i}{16}X_IF^I_{\nu\rho}\left(\Gamma_\mu\,^{\nu\rho}-6\delta_\mu^\nu\Gamma^\rho\right)\right]\epsilon\,, \notag \\
0\,=&\,\left[-\frac{i}{2}\mathcal{G}_{ij}\partial_\mu\varphi^j\Gamma^\mu+\frac{i}{2}\partial_iW+\frac{1}{4}\partial_iX_IF^I_{\mu\nu}\Gamma^{\mu\nu}\right]\epsilon\,,
\end{align}
where $\Gamma_\mu$ are Cliff(1,5) generators. The superpotential is given by
\begin{equation}
W\,\equiv\,\xi_IX^I\,,
\end{equation}
and the scalar potential can be expressed in terms of the superpotential,
\begin{equation} \label{potsup}
\mathcal{V}\,=\,\mathcal{G}^{ij}\partial_iW\partial_jW-\frac{5}{4}W^2\,.
\end{equation}

The supersymmetric $AdS_6$ vacuum of the scalar potential, which uplifts to the $AdS_6\times\widetilde{S}^4$ solution, \cite{Brandhuber:1999np}, is dual to a class of five-dimensional superconformal field theories, \cite{Seiberg:1996bd}, and the free energy is, \cite{Jafferis:2012iv},
\begin{equation}
\mathcal{F}_{S^5}\,=\,-\frac{\pi^2}{3}\frac{1}{G_N^{(6)}}\,=\,-\frac{9\sqrt{2}\pi}{5}\frac{N^{5/2}}{\sqrt{8-N_f}}\,.
\end{equation}

\subsection{The $AdS_4$ ansatz}

We consider the following background, \cite{BenettiGenolini:2024kyy},
\begin{equation} \label{ads4m2met}
ds_6^2\,=\,e^{2\lambda}\left[ds_{AdS_4}^2+ds_{M_2}^2\right]\,,
\end{equation}
where $ds^2_{AdS_4}$ is a metric on $AdS_4$ of unit radius.  The scalar fields and $\lambda$ are functions on $M_2$ and $A^I$ are gauge fields on $M_2$. 

We dimensionally reduce the equations of motion from the action in \eqref{sixtheory} on the background, \eqref{ads4m2met}, and find that the reduced equations of motion can be obtained from the variations of following two-dimensional action, $S_6=\frac{\text{vol}_{AdS_4}}{16\pi{G}_N^{(6)}}S_2$,
\begin{align} \label{2daction}
S_2\,=\,\left.\int_{M_2}\right[&e^{4\lambda}\left(R-12+20\left(\nabla\lambda\right)^2-\mathcal{G}_{ij}\partial_\mu\varphi^i\partial^\mu\varphi^j\right)-e^{6\lambda}\mathcal{V} \notag \\ 
&-\left.\frac{1}{2}e^{2\lambda}G_{IJ}F_{\mu\nu}^IF^{J\mu\nu}\right]\text{vol}\,,
\end{align}
where $R$, $\nabla$ and vol are the Ricci scalar, Levi-Civita connection and volume form on $M_2$. The Maxwell equations are
\begin{equation} \label{maxwell}
d\left[e^{2\lambda}G_{IJ}\left(*F^J\right)\right]\,=\,0 \qquad \Longrightarrow \qquad d\left[e^{2\lambda}G_{IJ}F_{12}^J\right]\,=\,0\,, \qquad \forall I\,=\,0,1,2\,,
\end{equation}
where we have $F^I\,=\,F^I_{12}\text{vol}$ with $F_{12}^I$ a function on $M_2$. For the warp factor, $\lambda$, the equation of motion is
\begin{align} \label{warpeom}
e^{4\lambda}\left[R-12+20\left(\nabla\lambda\right)^2-\mathcal{G}_{ij}\partial_\mu\varphi^i\partial^\mu\varphi^j\right]\,=&\,\frac{3}{2}e^{6\lambda}\mathcal{V}+\frac{1}{4}e^{2\lambda}G_{IJ}F_{\mu\nu}^IF^{J\mu\nu} \notag \\
&+\text{total\,\,derivative}\,.
\end{align}
The trace of Einstein equation is
\begin{equation} \label{traceeinstein}
e^{6\lambda}\mathcal{V}-\frac{1}{2}e^{2\lambda}G_{IJ}F_{\mu\nu}^{I}F^{J\mu\nu}\,=\,-12e^{4\lambda}+\text{total\,\,derivative}\,.
\end{equation}

From \eqref{2daction} the partially off-shell (POS) action is obtained by imposing \eqref{warpeom},
\begin{equation}
S_2|_\text{POS}\,=\,\frac{1}{2}\int_{M_2}\left(e^{6\lambda}\mathcal{V}\text{vol}-e^{2\lambda}G_{IJ}F_{12}^IF^J\right)\,.
\end{equation}
Further imposing \eqref{traceeinstein} we find the on-shell action,
\begin{equation}
S_2|_\text{OS}\,=\,-6\int_{M_2}e^{4\lambda}\text{vol}\,.
\end{equation}
The free energy of dual 3d SCFT. We introduce the trial free energy for $F$-maximization,
\begin{equation} \label{finalfree}
F^-\,=\,\left(\frac{2\pi^2}{81G_N^{(6)}}g^3m\right)S_2|_\text{POS}\,=\,-\left(\frac{2}{27}g^3m\right)\mathcal{F}_{S^5}S_2|_\text{POS}\,,
\end{equation}

\subsection{Equivariantly closed forms}

We introduce the Killing spinor on the background, \eqref{ads4m2met}, for the supersymmetry variations, \eqref{susyvar11}, \cite{BenettiGenolini:2024kyy},
\begin{equation}
\epsilon\,=\,\vartheta\otimes{e}^{\lambda/2}\zeta\,,
\end{equation}
where $\vartheta$ is a Killing spinor on $AdS_4$, $\zeta$ is a spinor on $M_2$ and the warp factor is introduced for convenience. Then, as presented in appendix \ref{appB}, we obtain the supersymmetry equations, \eqref{kse11}.

We introduce the spinor bilinears in $\zeta$ on $M_2$, \cite{BenettiGenolini:2024kyy},
\begin{equation} \label{sbil}
S\,=\,\zeta^\dagger\zeta\,, \quad P\,=\,\zeta^\dagger\gamma_3\zeta\,, \quad K\,=\,\zeta^\dagger\gamma_{(1)}\zeta\,, \quad \xi^\flat\,=\,-i\zeta^\dagger\gamma_{(1)}\gamma_3\zeta\,,
\end{equation}
where we have $\gamma_{(n)}=\frac{1}{n!}\gamma_{\mu_1\cdots\mu_n}dx^{\mu_1}\wedge\cdots\wedge{d}x^{\mu_n}$ and the chirality operator on $M_2$ is $\gamma_3=-i\gamma_1\gamma_2$. We obtain the spinor bilinear equations in \eqref{sbileq}. We find that the Killing vector, $\xi$, on $M_2$ is dual to the one-form, $\xi^\flat$, and we have
\begin{equation}
d\xi^\flat\,=\,-2\left(3+PS^{-1}e^\lambda{W}\right)P\text{vol}\,.
\end{equation}

We found equivariantly closed forms under $d_\xi=d-\xi\mathbin{\lrcorner}$ in appendix \ref{appB}, \cite{BenettiGenolini:2024kyy},
\begin{equation} \label{phiF}
\Phi^{F^I}\,\equiv\,F^I-X^Ie^\lambda{P}\,,
\end{equation}
and
\begin{equation}
\Phi^\text{vol}\,\equiv\,e^{6\lambda}\mathcal{V}\text{vol}-e^{5\lambda}WS\,.
\end{equation}
By employing the Maxwell equations in \eqref{maxwell}, we obtain another equivariantly closed form,
\begin{align} \label{phiS}
\Phi^S\,\equiv&\,\frac{1}{2}\left(\Phi^\text{vol}-e^{2\lambda}{G}_{IJ}F^I_{12}\Phi^{F^I}\right) \notag \\
=&\,\frac{1}{2}\left(e^{6\lambda}\mathcal{V}\text{vol}-e^{2\lambda}G_{IJ}F^I_{12}F^J-e^{5\lambda}WS+e^{3\lambda}F^I_{12}X^JP\right)\,,
\end{align}
where the two-forms are the two-dimensional action in \eqref{2daction}.

As we note that the scalar bilinear, $S$, is a constant in appendix \ref{appB}, we choose the normalization to be $S\equiv1$.

\subsection{Localization}

For the spinor bilinears in \eqref{sbil}, we find $||\xi||^2=||K||^2=S^2-P^2$. At a fixed point, $p$, where we have $\xi|_p=0$, we obtain $P|_p=\pm{S}|_p=\pm1$, and the sign is the chirality of the spinor, $\zeta$, at the fixed point, $p$. For $\xi\ne0$, around the fixed point, $p$, we may have $M_2$ to be $\mathbb{R}^2/\mathbb{Z}_{d_p}$ with orbifold singularities and $\xi=\epsilon_p\partial_{\phi_p}$ where $\phi_p$ is a polar coordinate of period, $2\pi/d_p$. Let us name $\epsilon_p$ to be the weight of $\xi$ at the fixed point, $p$, \cite{BenettiGenolini:2024kyy}.

We employ the Berline-Vergne-Atiyah-Bott fixed point formula, \cite{Berline:1982a, Atiyah:1984px}, to perform the integrations of equivariantly closed forms in \eqref{phiF} and \eqref{phiS} on $M_2$, \cite{BenettiGenolini:2024kyy}. The magnetic charges are given by
\begin{equation} \label{magch}
\mathfrak{p}^I\,\equiv\,\int_{M_2}\frac{F^I}{2\pi}\,=\,\int_{M_2}\frac{\Phi^{F^I}}{2\pi}\,=\,-\frac{1}{2\pi}\sum_{\text{fixed}\,p}\frac{1}{d_p}\frac{2\pi}{\epsilon_p}\left(X^Ie^\lambda{P}\right)\Big|_p\,,
\end{equation}
and the two-dimensional action is 
\begin{align} \label{twoac}
S_2|_\text{POS}\,=\,\int_{M_2}\Phi^S\,=&\,-\frac{1}{2}\sum_{\text{fixed}\,p}\frac{1}{d_p}\frac{2\pi}{\epsilon_p}\left(e^{5\lambda}WS-e^{3\lambda}G_{IJ}F_{12}^IX^JP\right)\Big|_p \notag \\
=&\,2\sum_{\text{fixed}\,p}\frac{1}{d_p}\frac{2\pi}{\epsilon_p}e^{4\lambda}P\Big|_p\,,
\end{align}
where we employed the algebraic constraint in \eqref{algecon} in the last line and $P^2|_p=S^2|_p=1$. As we have $d\xi^\flat|_p=2\epsilon_p\text{vol}$ at a fixed point, we find
\begin{equation} \label{dxidxi}
e^\lambda{W}P|_p\,=\,-\left(3+\epsilon_pP_p\right)\,,
\end{equation}
where we used $P_p\,=\,\pm1$.

We introduce the new scalar fields, 
\begin{equation} \label{newsc}
x^I\,\equiv\,-X^Ie^\lambda{P} \qquad \Longrightarrow \qquad \Phi^{F^I}\,=\,F^I+x^I\,.
\end{equation}
With $\mathcal{F}\left(X^I\right)=1$ in \eqref{prepotential}, we obtain
\begin{equation}
\mathcal{F}\left(x^I\right)|_p\,=\,x^0\left(x^1x^2\right)^{3/2}|_p\,=\,-e^{4\lambda}|_p\,,
\end{equation}
where we employed $P^2|_p=1$. Then \eqref{twoac}, \eqref{magch} and \eqref{dxidxi} are given, respectively, by
\begin{equation} \label{constraint11}
S_2|_\text{POS}\,=\,-2\sum_{\text{fixed}\,p}\frac{2\pi}{d_p\epsilon_p}\mathcal{F}\left(x_p^I\right)P_p\,, \qquad \mathfrak{p}^I\,=\,\sum_{\text{fixed}\,p}\frac{1}{d_p\epsilon_p}x_p^I\,, \qquad \xi_Ix_p^I\,=\,3+\epsilon_pP_p\,,
\end{equation}
where we introduced $x_p^I\equiv{x}^I|_p$.

Topologically a spindle is the only choice for $M_2$ if it is compact without boundary and has a $U(1)$ isometry with fixed points, \cite{Faedo:2021nub}. The spindle has two fixed points, $p=\pm$, and we choose $d_\pm=n_\pm$ with $n_\pm\ge1$. The Killing vector field on the spindle is given by
\begin{equation}
\xi\,=\,b_0\partial_\varphi\,,
\end{equation}
where $\varphi$ is an azimuthal coordinate on the spindle with period of $2\pi$. We can write $\epsilon_+=-b_0/n_+$, $\epsilon_-=b_0/n_-$. $P_\pm$ is arbitrary and $|P_\pm|=1$. Then the constraints in \eqref{constraint11} are given by
\begin{equation} \label{constraints22}
\mathfrak{p}^I\,=\,-\frac{1}{b_0}\left(x_+^I-x_-^I\right)\,, \qquad \xi_Ix_+^I\,=\,3-\frac{b_0}{n_+}P_+\,, \qquad \xi_Ix_-^I\,=\,3+\frac{b_0}{n_-}P_-\,,
\end{equation}
and, from these, we also obtain
\begin{equation} \label{magneticcon}
\xi_I\mathfrak{p}^I\,=\,\frac{P_+}{n_+}+\frac{P_-}{n_-}\,=\,\frac{n_-P_++n_+P_-}{n_-n_+}\,.
\end{equation}
If we introduce $\sigma\equiv{P}_+/P_-$, $\sigma=+1$ is for the case of equal chirality of spinor at two poles of spindle, namely twist. $\sigma=-1$ is for the opposite chiralities, namely anti-twist,\cite{Ferrero:2021etw}. The off-shell action is given by
\begin{equation} \label{S2pos22}
S_2|_\text{POS}\,=\,\frac{4\pi}{b_0}\left[\mathcal{F}\left(x_+^I\right)P_+-\mathcal{F}\left(x_-^I\right)P_-\right]\,,
\end{equation}
and we find the final expression of the off-shell free energy from \eqref{finalfree}.


\subsection{The free energy}


We introduce a new parametrization for the scalar fields with magnetic charges,
\begin{equation} \label{phi012}
\phi_0\,\equiv\,m\left(x_\pm^0\pm\mathfrak{p}^0\frac{b_0}{2}\right)\,, \qquad \phi_1\,\equiv\,g\left(x_\pm^1\pm\mathfrak{p}^1\frac{b_0}{2}\right)\,, \qquad \phi_2\,\equiv\,g\left(x_\pm^2\pm\mathfrak{p}^2\frac{b_0}{2}\right)\,.
\end{equation}
As the gauge field, $A^0$, is identically zero, \eqref{gaugefd}, we have $\mathfrak{p}^0=0$. Accordingly, we choose $\phi_0=1$. Namely, we have
\begin{equation}
\phi_0\,=\,1\,, \qquad \mathfrak{p}^0\,=\,0\,.
\end{equation}
For the choice of $\phi_0=1$, from \eqref{phi012} and \eqref{newsc}, we find 
\begin{equation}
1\,=\,m\left(-X^0e^\lambda{P}\right)|_\pm\,.
\end{equation}
As $X^0>0$ and $e^\lambda>0$, only the solutions in the twist class are allowed, $P_\pm=-1$, and it is consistent with \cite{Faedo:2021nub}. We further introduce new parameters for the rest of magnetic charges and $b_0$,
\begin{equation}
\epsilon\,\equiv\,\frac{b_0}{2}\,, \qquad \mathfrak{n}_1\,\equiv\,g\mathfrak{p}^1\,, \qquad \mathfrak{n}_2\,\equiv\,g\mathfrak{p}^2\,.
\end{equation}
Then, from the constraints in \eqref{constraints22}, using the new parameters, we find 
\begin{align} \label{gbcon}
\mathfrak{n}_1+\mathfrak{n}_2\,=&\,\frac{n_++\sigma{n}_-}{n_-n_+}\,, \notag \\
\phi_1+\phi_2-\frac{n_+-\sigma{n}_-}{n_-n_+}\epsilon\,=&\,2\,,
\end{align}
where we keep $\sigma\equiv{P}_+/P_-$ arbitrary. These are precisely the constraints on the magnetic charges and the variables, $\phi_i$, for D4-branes wrapped on a spindle in section 5.3 of \cite{Faedo:2021nub}.{\footnote{In \cite{Faedo:2021nub} $\phi_i$ was denoted by $\varphi_i$.}} The prepotential is then given by
\begin{equation}
\mathcal{F}\left(x_\pm^I\right)\,=\,\frac{1}{g^3m}\Big(\left(\phi_1\mp\mathfrak{n}_1\epsilon\right)\left(\phi_2\mp\mathfrak{n}_2\epsilon\right)\Big)^{3/2}\,.
\end{equation}
From \eqref{S2pos22}, the two-dimensional off-shell action is 
\begin{equation}
S_2|_\text{POS}\,=\,-\frac{2\pi}{\epsilon}\frac{1}{g^3m}\left[\Big(\left(\phi_1+\mathfrak{n}_1\epsilon\right)\left(\phi_2+\mathfrak{n}_2\epsilon\right)\Big)^{3/2}-\Big(\left(\phi_1-\mathfrak{n}_1\epsilon\right)\left(\phi_2-\mathfrak{n}_2\epsilon\right)\Big)^{3/2}\right]\,.
\end{equation}
From this, we recover the gravitational block in \cite{Faedo:2021nub},
\begin{equation} \label{offfree}
F^-\left(\phi_i,\epsilon;\mathfrak{n}_i\right)\,=\,\left(\frac{2\sqrt{2}}{15}\frac{N^{5/2}}{\sqrt{8-N_f}}g^3m\right)S_2|_\text{POS}\,.
\end{equation}
By extremizing the gravitational block or the off-shell action subject to the constraints, \eqref{gbcon},  we find, \cite{Faedo:2021nub},
\begin{align} \label{phistar}
\phi_1^*\,=&\,1+\frac{2x\left[\left(n_+-\sigma{n}_-\right)x+n_+n_-\left(\mathfrak{n}_1-\mathfrak{n}_2\right)\right]}{\left(x^2+3\right)\left[\left(n_+-\sigma{n}_-\right)-\left(n_++\sigma{n}_-\right)x\right]}\,, \notag \\
\epsilon^*\,=&\,\frac{4n_+n_-x^2}{\left(x^2+3\right)\left[\left(n_+-\sigma{n}_-\right)-\left(n_++\sigma{n}_-\right)x\right]}\,,
\end{align}
where $x$ is the only solution in the interval $(0,1)$ of the quartic equation,
\begin{align}
\left(n_++\sigma{n}_-\right)^2x^4+4&\left[2n_+^2n_-^2\left(\mathfrak{n}_1-\mathfrak{n}_2\right)^2-3\left(n_+^2+n_-^2-\sigma{n}_+n_-\right)\right]x^2 \notag \\
+12\left(n_+^2-n_-^2\right)&x-9\left(n_+-\sigma{n}_1\right)^2\,=\,0\,.
\end{align}
Inserting $\phi_1^*$ and $\epsilon^*$ in \eqref{offfree}, we obtain the free energy of dual three-dimensional superconformal field theories, \cite{Faedo:2021nub},
\begin{equation}
F^-\left(\phi_i^*,\epsilon^*;\mathfrak{n}_i\right)\,=\,\frac{\sqrt{3}\pi}{5}\frac{N^{5/2}}{\sqrt{8-N_f}}\frac{\left[3\left(n_+-\sigma{n}_-\right)\left(x^2+1\right)-\left(n_++\sigma{n}_-\right)x\left(x^2+5\right)\right]^{3/2}}{n_+n_-x\left(x^2+3\right)\left[\left(n_+-\sigma{n}_-\right)-\left(n_++\sigma{n}_-\right)x\right]^{1/2}}\,,
\end{equation}
for the solutions in the twist class, $\sigma=+1$.

By plugging $x_-^I=x_+^I+b_0\mathfrak{p}^I$ from \eqref{constraints22} into \eqref{S2pos22} with $\sigma=+1$, we note that
\begin{equation} \label{spherelim}
\lim_{b_0\rightarrow0}S_2|_\text{POS}\,=\,-4\pi\mathfrak{p}^I\frac{\partial}{\partial{x}^I}\mathcal{F}\left(x^I\right)\,.
\end{equation}
We have $x_-^I=x_+^I=x^I$ to be constant on $M_2$. From \eqref{magneticcon} we find $\xi_Ix^I=2$. We can extremize \eqref{spherelim} with the constraint, $\xi_Ix^I=2$. In particular, for $n_+=n_-=1$, we should find the extremization performed in \cite{Hosseini:2018usu} which calculates the holographic free energy of $AdS_4\times\Sigma_\mathfrak{g}$ solutions with a topological twist on the Riemann surface of genus, $\mathfrak{g}$, \cite{Nunez:2001pt, Naka:2002jz, Karndumri:2015eta, Bah:2018lyv, Hosseini:2018usu}, and we have $\xi_I\mathfrak{p}^I=2P_+$ and $\sigma=+1$.

\section{Conclusions} \label{sec4}

In this work, following the work of \cite{BenettiGenolini:2024kyy}, in six- and seven-dimensional gauged supergravity coupled to a vector multiplet on an ansatz of $AdS_{5,4}\times{M}_2$, respectively, where $M_2$ is a two-dimensional surface, we have constructed equivariantly closed forms from the spinor bilinears of Killing spinors. From the integration of equivariantly closed forms via the Berline-Vergne-Atiyah-Bott fixed point formula, we derived the gravitational blocks which were previously conjectured in \cite{Faedo:2021nub}. For the construction of equivariantly closed forms, we formulated six- and seven-dimensional gauged supergravity theories in the structure of very special geometry.

In order to formulate six- and seven-dimensional gauged supergravity in the very special geometry, it appears to require the introduction of a trivial gauge field, $A^0$, in $A^I\,=\,\left(A^0\equiv0,A^1,A^2\right)$ as in \eqref{gaugefd7} and \eqref{gaugefd}. It would be nice to better understand the structure of special geometry in six and seven dimensions.

The common structure of very special geometry and the similarity of formulations of gauged supergravity in six and seven dimensions here and also in five dimensions in \cite{BenettiGenolini:2024kyy} are striking. It may hint a universal structure of gauged supergravity in diverse dimensions which can be utilized to construct solutions and study the geometry of theories. Thus, the study of very special geometry in six- and seven-dimensional gauged supergravity coupled to an arbitrary number of vector multiplets is requested. It would be also interesting to see if there is a structure of very special geometry in four-dimensional theories as well.

A natural extension of the study is to consider the equivariant localization for M5-branes and D4-branes wrapped on four-dimensional manifolds, including orbifolds, \cite{Giri:2021xta, Faedo:2021nub, Suh:2022olh, Cheung:2022ilc, Couzens:2022lvg, Faedo:2022rqx, Faedo:2024upq}.

\bigskip
\bigskip
\leftline{\bf Acknowledgements}
\noindent This work was supported by the National Research Foundation of Korea (NRF) grant
funded by the Korea government, Ministry of Science and ICT (RS-2026-25481139).

\appendix
\section{Reduction of Killing spinors in seven dimensions} \label{appA}
\renewcommand{\theequation}{A.\arabic{equation}}
\setcounter{equation}{0} 

In this appendix we reduce the supersymmetry variations, \eqref{susyvar117}, on the background of $AdS_5\times{M}_2$, \eqref{ads4m2met7}, by following \cite{BenettiGenolini:2024kyy}.

\subsection{Killing spinor equations} \label{appA11}

We employ the spinor and gamma matrices, \cite{BenettiGenolini:2024kyy},
\begin{equation}
\epsilon\,=\,\vartheta\otimes{e}^{\lambda/2}\zeta\,, \quad \Gamma_\mathtt{i}\,=\,\beta_\mathtt{i}\otimes\gamma_3\,, \quad \Gamma_{a+2}\,=\,{1}\otimes\gamma_a\,,
\end{equation}
where we have frame indices on $AdS_5$, $\mathtt{i}=0,1,\ldots,4$, and on $M_2$, $a=1,2$ and introduce the chirality operator on $M_2$, $\gamma_3=-i\gamma_1\gamma_2$. The Killing spinor equation on $AdS_5$ is
\begin{equation}
\nabla_\mathtt{i}\vartheta\,=\,\frac{1}{2}\beta_\mathtt{i}\vartheta\,.
\end{equation}
Then the supersymmetry variations, \eqref{susyvar117}, reduce to
\begin{align} \label{kse117}
\nabla_a\zeta\,=&\,\left[\frac{1}{2}\left(1-e^{-\lambda}G_{IJ}X^IF_{12}^J\right)\gamma_a\gamma_3+\frac{i}{2}Q_a\right]\zeta\,, \notag \\
0\,=&\,\left[\slashed{\partial}\lambda+\frac{1}{5}e^\lambda{W}+\left(1-\frac{1}{5}e^{-\lambda}G_{IJ}X^IF_{12}^J\right)\gamma_3\right]\zeta\,, \notag \\
0\,=&\,\left[\mathcal{G}_{ij}\slashed{\partial}\varphi^j-e^\lambda\partial_iW+e^{-\lambda}G_{IJ}\partial_iX^IF_{12}^J\gamma_3\right]\zeta\,.
\end{align}
For an arbitrary generator, $\mathbb{A}$, on $M_2$, we find the spinor bilinear equations,
\begin{align} \label{sbileq7}
\nabla_a\left(\zeta^\dagger\mathbb{A}\zeta\right)\,=&\,\frac{1}{2}\left(1-e^{-\lambda}G_{IJ}X^IF_{12}^J\right)\zeta^\dagger\left[\mathbb{A},\gamma_a\gamma_3\right]_-\zeta\,, \notag \\
0\,=\,&\left(\partial_a\lambda\right)\zeta^\dagger\left[\mathbb{A},\gamma^a\right]_\pm\zeta+\frac{1}{5}\left(1\pm1\right)e^\lambda{W}\zeta^\dagger\mathbb{A}\zeta \notag \\
&+\left(1-\frac{1}{5}e^{-\lambda}G_{IJ}X^IF_{12}^J\right)\zeta^\dagger\left[\mathbb{A},\gamma_3\right]_\pm\zeta\,, \notag \\
0\,=&\,\mathcal{G}_{ij}\left(\partial_a\varphi^j\right)\zeta^\dagger\left[\mathbb{A},\gamma^a\right]_\pm\zeta-\left(1\pm1\right)e^\lambda\partial_i{W}\zeta^\dagger\mathbb{A}\zeta \notag \\
&+e^{-\lambda}G_{IJ}\partial_iX^IF_{12}^J\zeta^\dagger\left[\mathbb{A},\gamma_3\right]_\pm\zeta\,,
\end{align}
where $[\,,\,]_\pm$ are, respectively, commutator and anti-commutator.

\subsection{Bilinear equations} \label{appA12}

The spinor bilinears in \eqref{sbil7}, satisfy, \cite{BenettiGenolini:2024kyy},
\begin{equation}
\xi\mathbin{\lrcorner}\text{vol}\,=\,-K\,, \qquad F_{12}^IK\,=\,-\xi\mathbin{\lrcorner}F^I\,.
\end{equation}
Then the spinor bilinear equations, \eqref{sbileq}, reduce to differential equations, 
\begin{align} \label{diffeq7}
&dS\,=\,0\,, \qquad dK\,=\,0\,, \qquad d\xi^\flat\,=\,-2\left(4+e^\lambda{W}PS^{-1}\right)P\text{vol}\,, \notag \\
&d\left(e^\lambda{P}\right)\,=\,-\frac{4}{5}G_{JK}X^J\left(\xi\mathbin{\lrcorner}F^K\right)\,, \qquad d\left(e^{6\lambda}S\right)\,=\,\frac{6}{5}e^{7\lambda}W\left(\xi\mathbin{\lrcorner}\text{vol}\right)\,, \notag \\
&dX^I\,=\,-e^{-\lambda}\mathcal{G}^{ij}G_{JK}\partial_iX^I\partial_jX^J\left(\xi\mathbin{\lrcorner}F^K\right)P^{-1} \notag \\
& \qquad =\,-e^\lambda\mathcal{G}^{ij}\partial_iX^I\partial_jW\left(\xi\mathbin{\lrcorner}\text{vol}\right)S^{-1}\,,
\end{align}
and an algebraic equation,
\begin{equation} \label{algecon7}
e^{-\lambda}X_IF^I_{12}\,=\,5+e^\lambda{W}PS^{-1}\,.
\end{equation}
Employing \eqref{diffeq7} and \eqref{GIJGIJ7}, we find
\begin{align}
d\left(X^Ie^\lambda{P}\right)\,=&\,\left(dX^I\right)e^\lambda{P}+X^Id\left(e^\lambda{P}\right) \notag \\
=&\,-G_{JK}\left(\mathcal{G}^{ij}\partial_iX^I\partial_jX^J+\frac{4}{5}X^IX^J\right)\left(\xi\mathbin{\lrcorner}F^K\right) \notag \\
=&\,-G_{JK}G^{IJ}\left(\xi\mathbin{\lrcorner}F^K\right)\,=\,-\xi\mathbin{\lrcorner}F^I\,,
\end{align}
and the poly-form, $\Phi^{F^I}\,=\,F^I-X^Ie^\lambda{P}$, is equivariantly closed. Employing \eqref{diffeq7} and \eqref{potsup7}, we find
\begin{align}
d\left(e^{6\lambda}WS\right)\,=&\,\xi_I\left(dX^I\right)e^{6\lambda}S+Wd\left(e^{6\lambda}S\right) \notag \\
=&\,-e^{7\lambda}\left(\mathcal{G}^{ij}\partial_iW\partial_jW-\frac{6}{5}W^2\right)\left(\xi\mathbin{\lrcorner}\text{vol}\right) \notag \\
=&\,-e^{7\lambda}\mathcal{V}\left(\xi\mathbin{\lrcorner}\text{vol}\right)\,,
\end{align}
and the poly-form, $\Phi^\text{vol}\,=\,e^{7\lambda}\mathcal{V}\text{vol}-e^{6\lambda}WS$, is equivariantly closed.

\section{Reduction of Killing spinors in six dimensions} \label{appB}
\renewcommand{\theequation}{B.\arabic{equation}}
\setcounter{equation}{0} 

In this appendix we reduce the supersymmetry variations, \eqref{susyvar11}, on the background of $AdS_4\times{M}_2$, \eqref{ads4m2met}, by following \cite{BenettiGenolini:2024kyy}.

\subsection{Killing spinor equations} \label{appB11}

We employ the spinor and gamma matrices, \cite{BenettiGenolini:2024kyy},
\begin{equation}
\epsilon\,=\,\vartheta\otimes{e}^{\lambda/2}\zeta\,, \quad \Gamma_\mathtt{i}\,=\,\beta_\mathtt{i}\otimes\gamma_3\,, \quad \Gamma_{a+2}\,=\,{1}\otimes\gamma_a\,,
\end{equation}
where we have frame indices on $AdS_4$, $\mathtt{i}=0,1,2,3$, and on $M_2$, $a=1,2$ and introduce the chirality operator on $M_2$, $\gamma_3=-i\gamma_1\gamma_2$. The Killing spinor equation on $AdS_4$ is
\begin{equation}
\nabla_\mathtt{i}\vartheta\,=\,\frac{1}{2}\beta_\mathtt{i}\vartheta\,.
\end{equation}
Then the supersymmetry variations, \eqref{susyvar11}, reduce to
\begin{align} \label{kse11}
\nabla_a\zeta\,=&\,\left[\frac{1}{2}\left(1-e^{-\lambda}G_{IJ}X^IF_{12}^J\right)\gamma_a\gamma_3+\frac{i}{2}Q_a\right]\zeta\,, \notag \\
0\,=&\,\left[\slashed{\partial}\lambda+\frac{1}{4}e^\lambda{W}+\left(1-\frac{1}{4}e^{-\lambda}G_{IJ}X^IF_{12}^J\right)\gamma_3\right]\zeta\,, \notag \\
0\,=&\,\left[\mathcal{G}_{ij}\slashed{\partial}\varphi^j-e^\lambda\partial_iW+e^{-\lambda}G_{IJ}\partial_iX^IF_{12}^J\gamma_3\right]\zeta\,.
\end{align}
For an arbitrary generator, $\mathbb{A}$, on $M_2$, we find the spinor bilinear equations,
\begin{align} \label{sbileq}
\nabla_a\left(\zeta^\dagger\mathbb{A}\zeta\right)\,=&\,\frac{1}{2}\left(1-e^{-\lambda}G_{IJ}X^IF_{12}^J\right)\zeta^\dagger\left[\mathbb{A},\gamma_a\gamma_3\right]_-\zeta\,, \notag \\
0\,=\,&\left(\partial_a\lambda\right)\zeta^\dagger\left[\mathbb{A},\gamma^a\right]_\pm\zeta+\frac{1}{4}\left(1\pm1\right)e^\lambda{W}\zeta^\dagger\mathbb{A}\zeta \notag \\
&+\left(1-\frac{1}{4}e^{-\lambda}G_{IJ}X^IF_{12}^J\right)\zeta^\dagger\left[\mathbb{A},\gamma_3\right]_\pm\zeta\,, \notag \\
0\,=&\,\mathcal{G}_{ij}\left(\partial_a\varphi^j\right)\zeta^\dagger\left[\mathbb{A},\gamma^a\right]_\pm\zeta-\left(1\pm1\right)e^\lambda\partial_i{W}\zeta^\dagger\mathbb{A}\zeta \notag \\
&+e^{-\lambda}G_{IJ}\partial_iX^IF_{12}^J\zeta^\dagger\left[\mathbb{A},\gamma_3\right]_\pm\zeta\,,
\end{align}
where $[\,,\,]_\pm$ are, respectively, commutator and anti-commutator.

\subsection{Bilinear equations} \label{appB12}

The spinor bilinears in \eqref{sbil}, satisfy, \cite{BenettiGenolini:2024kyy},
\begin{equation}
\xi\mathbin{\lrcorner}\text{vol}\,=\,-K\,, \qquad F_{12}^IK\,=\,-\xi\mathbin{\lrcorner}F^I\,.
\end{equation}
Then the spinor bilinear equations, \eqref{sbileq}, reduce to differential equations, 
\begin{align} \label{diffeq}
&dS\,=\,0\,, \qquad dK\,=\,0\,, \qquad d\xi^\flat\,=\,-2\left(3+e^\lambda{W}PS^{-1}\right)P\text{vol}\,, \notag \\
&d\left(e^\lambda{P}\right)\,=\,-\frac{3}{4}G_{JK}X^J\left(\xi\mathbin{\lrcorner}F^K\right)\,, \qquad d\left(e^{5\lambda}S\right)\,=\,\frac{5}{4}e^{6\lambda}W\left(\xi\mathbin{\lrcorner}\text{vol}\right)\,, \notag \\
&dX^I\,=\,-e^{-\lambda}\mathcal{G}^{ij}G_{JK}\partial_iX^I\partial_jX^J\left(\xi\mathbin{\lrcorner}F^K\right)P^{-1} \notag \\
& \qquad =\,-e^\lambda\mathcal{G}^{ij}\partial_iX^I\partial_jW\left(\xi\mathbin{\lrcorner}\text{vol}\right)S^{-1}\,,
\end{align}
and an algebraic equation,
\begin{equation} \label{algecon}
e^{-\lambda}X_IF^I_{12}\,=\,4+e^\lambda{W}PS^{-1}\,.
\end{equation}
Employing \eqref{diffeq} and \eqref{GIJGIJ}, we find
\begin{align}
d\left(X^Ie^\lambda{P}\right)\,=&\,\left(dX^I\right)e^\lambda{P}+X^Id\left(e^\lambda{P}\right) \notag \\
=&\,-G_{JK}\left(\mathcal{G}^{ij}\partial_iX^I\partial_jX^J+\frac{3}{4}X^IX^J\right)\left(\xi\mathbin{\lrcorner}F^K\right) \notag \\
=&\,-G_{JK}G^{IJ}\left(\xi\mathbin{\lrcorner}F^K\right)\,=\,-\xi\mathbin{\lrcorner}F^I\,,
\end{align}
and the poly-form, $\Phi^{F^I}\,=\,F^I-X^Ie^\lambda{P}$, is equivariantly closed. Employing \eqref{diffeq} and \eqref{potsup}, we find
\begin{align}
d\left(e^{5\lambda}WS\right)\,=&\,\xi_I\left(dX^I\right)e^{5\lambda}S+Wd\left(e^{5\lambda}S\right) \notag \\
=&\,-e^{6\lambda}\left(\mathcal{G}^{ij}\partial_iW\partial_jW-\frac{5}{4}W^2\right)\left(\xi\mathbin{\lrcorner}\text{vol}\right) \notag \\
=&\,-e^{6\lambda}\mathcal{V}\left(\xi\mathbin{\lrcorner}\text{vol}\right)\,,
\end{align}
and the poly-form, $\Phi^\text{vol}\,=\,e^{6\lambda}\mathcal{V}\text{vol}-e^{5\lambda}WS$, is equivariantly closed.

\vspace{2cm}

\bibliographystyle{JHEP}
\bibliography{20240307_ref}

\end{document}